\documentstyle[11pt,epsf]{article}

\input psfig
\def\beq{\begin{eqnarray}}
\def\eeq{\end{eqnarray}}
\def\bea{\begin{eqnarray*}}
\def\eea{\end{eqnarray*}}

\def\centeron#1#2{{\setbox0=\hbox{#1}\setbox1=\hbox{#2}\ifdim
\wd1>\wd0\kern.5\wd1\kern-.5\wd0\fi
\copy0\kern-.5\wd0\kern-.5\wd1\copy1\ifdim\wd0>\wd1
\kern.5\wd0\kern-.5\wd1\fi}}
\def\ltap{\;\centeron{\raise.35ex\hbox{$<$}}{\lower.65ex\hbox{$\sim$}}\;}
\def\gtap{\;\centeron{\raise.35ex\hbox{$>$}}{\lower.65ex\hbox{$\sim$}}\;}

\def\singleandthirdspaced{\baselineskip=\normalbaselineskip\multiply
    \baselineskip by 130\divide\baselineskip by 100}
\def\singlespaced{\baselineskip=\normalbaselineskip}

\newcommand{\newc}{\newcommand}
\newc{\qbar}{{\overline q}}
\newc{\Kahler}{K\"ahler }
\newc{\deltaGS}{\delta_{\rm GS}}
\begin{document}
\begin{titlepage}
\begin{flushright}
{\large hep-th/0001157 \\ SCIPP-99/22\\
}
\end{flushright}

\vskip 1.2cm

\begin{center}

{\LARGE\bf Some Reflections on Moduli, Their Stabilization and
Cosmology}

\vskip 1.4cm

{\large  Michael Dine\footnote{invited talk at ITP
Conference on New Dimensions in Field Theory and
String Theory, ITP Nov17, 1999}}
\\
\vskip 0.4cm
{\it Santa Cruz Institute for Particle Physics,
     Santa Cruz CA 95064  } \\

\vskip 4pt

\vskip 1.5cm

\begin{abstract}
We review some aspects of moduli in string theory.  We argue that
one should focus on {\it approximate moduli spaces}, and that
there is evidence that such spaces exist non-perturbatively.  We
ask what it would mean for string theory to predict low energy
supersymmetry.  Aspects of two proposed mechanisms for fixing the
moduli are discussed, and solutions to certain cosmological
problems associated with moduli are proposed.

\end{abstract}

\end{center}

\vskip 1.0 cm

\end{titlepage}
\setcounter{footnote}{0} \setcounter{page}{2}
\setcounter{section}{0} \setcounter{subsection}{0}
\setcounter{subsubsection}{0}

\singleandthirdspaced


\section{Introduction}

Moduli are a crucial ingredient in all of our present
understanding of string theory, both at a perturbative and a
non-perturbative level.  Such particles are not present in nature,
and, since string theory otherwise looks a lot like nature, most
of us have adopted the view that these moduli will somehow be
fixed, or perhaps, in the final description, won't be there at
all. Susskind has put forth a provocative, contrary view.  He
asks:  what is strong theory?, and answers:  it is that theory
which lives on those moduli spaces with enough supersymmetry that
we can make definite statements about it, not only perturbatively
but non-perturbatively. Such non-perturbative statements include
understandings from dualities and, in some cases, non-perturbative
formulations (Matrix Theory and the AdS correspondence) at least
in some regions of the moduli space (Matrix theory and the AdS
correspondence).

This definition goes against what has
become the almost conventional wisdom that any quantum theory of
gravity is string theory; it is provocative precisely because it excludes the
world we see.  But it is also provocative because it is
challenges us to decide: what would constitute a signature for string
theory.  One example which is often discussed is low energy
supersymmetry.  In various
approximations, low energy supersymmetry emerges naturally from
string theory, but it is not at all clear that it is a necessary
outcome.  This question is brought into sharp focus by
recent
proposals to solve the hierarchy problem\cite{largedimensions,precursors,rs},
some within a string-theoretic
framework, which {\it do not} invoke low energy
supersymmetry.  In fact, in the spirit of Susskind's challenge, we can
first try to define what we might mean by such
a prediction.   Supersymmetry, after all, must be a broken symmetry,
so what we are asking is whether string theory predicts
supersymmetry broken ``a little bit," rather than
a lot.  While we might wish to argue for this based
on considerations of hierarchy, it would be far better if we could see that
such breaking was intrinsic to string theory.  Of course, we might hope to
simply solve for some ground state of string theory which has all
sorts of desired properties.  This state might have some low energy
supersymmetry, but (modulo one hopeful comment to appear below) it
doesn't seem likely that this will happen soon.    Instead, we might
try to ask if there is a sense in which low energy supersymmetry
might be preferred or generic.

I won't offer an answer to this question, but will try, instead,
to state the question in a precise way.  What would it mean to
predict low energy supersymmetry?  After all, how can we tell, without an
actual calculation, supersymmetry broken by a small amount from
supersymmetry broken by an amount of order one.  The key to
formulating this question is the notion of ``approximate moduli."
While such a notion falls outside of the strict definition
proposed by Susskind, I believe it is quite plausible.  Indeed, I
would offer two pieces of evidence to suggest we should enlarge
Susskind's definition of string theory.  First, there are string
vacua with $N=1$ supersymmetry which can be constructed
perturbatively, and for which one can show that a subspace of the
moduli space cannot be lifted non-perturbatively, provided that
the theory exists on this space. While one does not have a
non-perturbative definition of these states, the strong
constraints provided by symmetries and holomorphy are very
suggestive\cite{bdmoduli}.  Moreover, these states can be studied both
at strong and weak coupling, using Heterotic-Type I and
Heterotic-Heterotic string dualities.

Still more interesting, however, are theories where the moduli are
lifted.  Here one might worry, in the spirit
of Susskind's remarks, that the theories do not exist.
But consider the case of large radius compactifications of the
heterotic string on Calabi-Yau spaces.  At weak coupling the
construction is well-known.  Classically, there are several
moduli.  One, called $S$, determines the tree level gauge couplings;
another set, referred to as $T$, determine the size and shape of
the internal space.  We will caricature this slightly, and
call $T=R^2/\ell_{st}^2$, $S= g_{s}^{-2}V/\ell_{s}^6$ ($\ell_{s}$
is the string length, $\sqrt{\alpha^{\prime}}$).
In order that string perturbation theory be
valid, one needs
\beq
S \gg 1~~~~~~ S/ T^3 \gg 1.
\eeq
One also believes one knows how to describe these states at strong coupling,
\beq
S \gg 1 ~~~~~ T \gg 1 ~~~~~T^3/S \gg 1
.\eeq
Here an eleven dimensional description is appropriate\cite{hv}.
If $\rho$ is
the size of the eleventh dimension, and $R$ the size of the internal
space, $T$
now corresponds to $R^2 \rho/\ell_{11}^3$, while $S =
V/\ell_{11}^6$ ($\ell_{11}$ is the eleven dimensional Planck
length).

Using holomorphy arguments and the $2\pi$ periodicities of the
axions\cite{kahlerstabilization}, it follows that the gauge coupling
functions are of the form
\beq
f^a= S + n_a T + {\cal O}(e^{-S},e^{-T}).
\eeq
In both the strong and weak
coupling regimes, one can have $S \gg 1$, $T \gg 1$, so if the theory
is consistent, the $f^a$'s should agree in both the strong and
weak coupling regimes, and indeed they do\cite{wittency,bdcouplings}.

Generically, however, potentials are generated for the moduli in
these states.  Any superpotential due to short distance string effects
is proportional to $e^{-S}$, $e^{-T}$.
Low energy effects will generically generate a {\it far larger}
contribution to the superpotential.  Gluino condensation is
a well-known example, where one generates a superpotential of the
form
\beq
W= e^{(-S-nT)/3 b_o}.
\eeq
The corrections to this expression are generally of order ${\cal
O}(e^{-2(-S-nT)/3 b_o})$, so for suitably large $S$ and $T$, the
superpotential calculation should be reliable, both in the strong
and weak coupling regimes.
Again, if it makes sense to speak of these states,
these expressions must agree between the weak and strong
coupling regimes, and they do.

This agreement is strong evidence that these states exist, and
that, for large $S$ and $T$, it makes sense to speak of
{\it approximate moduli} and an an approximate moduli space.
In fact, even the Kahler potentials agree in this
region.  This can be understood from the fact that, for
appropriate $T$ and $S$, in both regimes, the theory is
approximately $10$-dimensional, with $N=1$ supersymmetry.  The
supersymmetry uniquely determines the form of the ten dimensional
lagrangian, up to higher derivative corrections, and this
lagrangian, in turn, determines the structure of the leading
Kahler potential terms.  In other words, there is good reason to
enlarge Susskind's definition at least to a large class of
approximate moduli spaces which have $N=1$ supersymmetry in
various limits.  Whether there is a similar class of $N=0$
theories is clearly a very important question.  As we will discuss
later, it is crucially related to the question:  does string
theory predict low energy supersymmetry.

In this talk, I will focus on the
question of stabilization.  In particular, I want to consider the
possibility that the true vacuum of string theory lies on such an
approximate moduli space.  This has been the viewpoint in
virtually all thinking about string phenomenology, both because it
is the best we can do at the moment, but also because it is
consistent with some basic facts of nature.  It need not hold.
But I wish to ask here two questions:
\begin{itemize}
\item
What sorts of stabilization mechanisms might explain this fact,
for particular approximate moduli spaces (note this replaces the
usual phrase, ``string vacuua").
I will focus particularly on the ``racetrack mechanism" as a
possible explanation\cite{racetrack,kl,iy,dsracetrack}.
\item
Can we phrase any the problem of obtaining
generic string predictions in this language.
\end{itemize}

Let me focus on the second question first, and consider, in
particular, the question of low energy supersymmetry as a string
theory prediction.  In the past, we have usually spoken about
``classical solutions which respect N=1 supersymmetry."  But we
are probably not interested in classical solutions.  So we need
a more precise definition. If
supersymmetry is broken, what distinguishes, say, the moduli
spaces of heterotic string compactifications on Calabi-Yau, from
those which, to leading order in coupling, have no supersymmetry?
The best formulation which I have been able to come up with is the
following.  We can distinguish two classes of moduli spaces
without supersymmetry:
\begin{itemize}
\item
Those in which supersymmetry is restored in certain regions of the
moduli space, and where there are only a finite number of light states
(compared to some characteristic energy scale)
in some of these regions.  In particular, in these regimes, there
is only one light spin 3/2 particle.  These we will refer to as approximate
string
moduli spaces with low energy supersymmetry, or LES.
\item
Those in which supersymmetry is restored only in regions where
there are an infinite number of light states.  In particular, in
these regimes, there are an infinite number of light spin 3/2
particles.   These we will describe as states with bulk susy.
In general, this bulk supersymmetry need not manifest itself in
low energy physics in any conventional sense.  Rather than light
spin-3/2 particles, there may be light spin-1/2 particles
associated with supersymmetry breaking by branes.  The
phenomenology of these is not generic.
\item
Moduli spaces in which supersymmetry is restored no where in the
moduli space.   We will refer to these simply as
non-supersymmetric strings.
\end{itemize}

Calabi-Yau spaces, both at weak and strong
coupling, are examples of the first.  As the heterotic
dilaton tends to infinity, supersymmetry is restored, and there
are only a finite number of states which are light compared
to the string scale.  One {\it can} take moduli to infinity
(the ``T" moduli) so as to obtain infinite numbers of states.  The second are
exemplified, for example, by Rohm type compactifications\cite{rohm}, where
light states only appear as some radius tends to infinity;  in
this limit, there are an infinite number of Kaluza-Klein states,
and in particular an infinite number of gravitinos.  Examples
of non-supersymmetric (approximate) moduli spaces are provided by
by compactification of the ten dimensional $O(16)\times
O(16)$ string.

Recent proposals for very large dimensions can be discussed in
this framework.  In particular, it has been argued that a low
string scale might explain the hierarchy without supersymmetry. At
the same time, most of these proposals invoke some degree of bulk
supersymmetry to resolve various questions, and in particular to
explain the size of the internal dimension without invoking large
numbers\cite{loghierarchies}.  One of the nicest proposals is that
supersymmetry is preserved in the bulk and on ``our" brane, while
being broken on some distant brane\cite{ibanez}. Such a proposal
is in the second class, even though its phenomenology resembles
that of conventional supergravity models.  (The detailed
phenomenology is different, since there is no scale at which the
conventional $N=1$ supergravity lagrangian is appropriate, so it
is probably possible, at least in principle, to distinguish these
possibilities.)  The third possibility might be illustrated by a
model of the Randall-Sundrum type.  The authors of
\cite{goldbergerwise} have argued that supersymmetry is not
necessary in this case to explain the large hierarchy.  Whether
this solution is really stable, for example, to perturbations of
the action used in these analyses (e.g. addition of ${\cal R}^2$
terms to the action) is a question which is currently under study,
though at the moment the answer seems to be
yes.\cite{graesseretal}.

This, then, is what one might mean by the statement that string
theory predicts low energy supersymmetry:  string theory predicts
that we sit in an approximate moduli space of the LES type.
Of course, we have not established that string theory makes
such a prediction, but at least we have succeeded (if somewhat
crudely) in stating what the question is.  There is certainly
a prejudice that the non-supersymmetric states may not exist.
They suffer tachyons and other instabilities, and it is hard
to see how one would set up any non-perturbative
formulation\cite{banksmotl}, but there is certainly no hard argument that
either the LES's make sense (and in particular, there is no
calculation which shows that the moduli are ever stabilized in
on such a space), or that the non-supersymmetric
states do not.

One, albeit limited, approach to this problem is the following.
One can go to points in these moduli spaces with discrete
symmetries, and ask whether or not these symmetries are anomalous.
Such an anomaly would almost certainly signal an inconsistency.
Searches among LES models have failed to yield such anomalies.
This appears non-trivial.  In many case, anomaly cancellation is
through a Green-Schwarz mechanism.  Currently, a search of moduli
spaces with bulk susy is in progress.  If one found examples of
theories with anomalies, this would suggest that, generically,
such models are inconsistent.

\section{Kahler Stabilization and The Racetrack Models}

In thinking about stabilization, there are some basic issues to
keep in mind.  Given the statement that string theory is a theory
without parameters, one would expect that stabilization should
occur in regimes where all of the couplings are strong.  There is nothing,
in principle, wrong with this possibility.
It is disappointing in
that it could mean that nothing is accessible to calculation,
and that any real predictions from string theory are beyond
reach.  On the
other hand, if we believe that string theory has anything to do
with nature, we need to explain the fact that the gauge couplings
are weak.  Indeed, this might make us optimistic that some
quantities, in the end, will be calculable.  Another source for
optimism is the existence of hierarchies:  this also suggests that
there should be small parameters in the problem.  Our earlier
discussion of approximate moduli spaces suggests that hierarchies
be explained by stabilization on such a space, where the
couplngs are small, and $e^{-S}$ and $e^{-T}$ are tiny.
Non-supersymmetric proposals involving large dimensions also
rely on such small factors\cite{loghierarchies,goldbergerwise}.
Of course, these facts could be accidents.  It could be, for
example, that a strong coupling theory accidentally predicts a
number of order $1/30$, and the hierarchy is the result of low
energy physics.  In this case, again, it would be hard to make
predictions from the underlying theory; indeed, it would be hard
to argue (in the spirit of Susskind's challenge) in what sense
this theory was connected with anything we call string theory.

The focus, then, of any attempt to understand the fate of the
moduli should be on the questions:
\begin{itemize}
\item Why are the gauge couplings small (and unified?)?
\item Why are there large hierarchies?
\item  Can anything be calculated, i.e. is it possible to relate
the phenomena to the underlying fundamental theory?
\end{itemize}

Within the framework of the LES models, there have been three
proposals for stabilization of the moduli.  The first is known as
``Kahler Stabilization"\cite{kahlerstabilization}."
Here one assumes that the moduli (say
$S$, the modulus which determines the gauge couplings) indeed are
large, so that holomorphic quantities are given by their forms in
the large $S$ limit (weak coupling string, or supergravity limit,
say); on the other hand, one supposes that the Kahler potential is
far from its weak coupling form, and is such that it leads to
stabilization.  In terms of the picture of the moduli space
which we have described for the heterotic string this seems
quite reasonable.  There are certainly regions of this moduli
space where holomorphic quantities are calculable, but
the Kahler potential is not.  It is possible, then, that the superpotential might be
understood as a result of gluino condensation, while the Kahler
potential would be quite different from its very weak
or very strong coupling form.  One can easily write down
Kahler potentials which stabilize the moduli, and, at the price
of a mysterious fine tuning, give vanishing cosmological
constant. If this picture is correct, holomorphic quantities can
be calculated by passing to one or the other limit, but
non-holomorphic quantities are not calculable.

The second scenario which has been widely considered in the
literature is known as the racetrack model\cite{racetrack,kl}.
The racetrack models illustrate several possibilities for stabilizing
the moduli:
\begin{itemize}
\item
Stabilization at scales of order one (with supersymmetry
broken or unbroken), and with nothing calculable.
\item
Stabilization, at the price of discrete fine tuning,
with a hierarchy of scales, and at least some quantities
calculable.
\end{itemize}

The racetrack idea is a very interesting proposal for
stabilization, which has existed for a number of years in a sort
of limbo, because of two issues.  First, there are no compelling
models in weakly coupled string theory for the phenomenon.
Second, it was hard to see why supersymmetry would be broken in a
suitable way (and certainly not with vanishing cosmological
constant).  Finally, and perhaps most fundamentally, given that
one was considering a theory without any parameters, how could
there really be a scheme in which one could hope to calculate
anything?

Kaplunovsky and Louis, however, revisited some of these issues a
few years ago, in light of duality and in particular in light of
developments connected with F theory\cite{kl}.  They pointed out that it is
easy to obtain huge groups in string theory, so constructing
models may not be such a problem.  They did not, however, really
address the question of what might be computable in such a scheme.
Their ideas, however, illustrate many of the main points.

In the racetrack model, one imagines one has two or more strongly
interacting gauge groups, and analyzes the problem from the
perspective of the low energy theory.  We have already argued that
such a low energy analysis is often appropriate.  Suppose, for
example, that one has several gauge groups in the low energy
theory without matter fields.  Then it is usually argued that the
superpotential has the form (in the notation of \cite{kl}:
\beq
W=  M_p^3 \sum_a C_a(T) e^{-{6\pi \over b_a \alpha_a(T)}}
.
\eeq
Kaplunovsky and Louis then argue that we now know
compactifications
for which the $b_a$'s are enormous, and that generic stationary
points of the potential may well have $6 \pi/\alpha \sim b_a$,
and that this is a way to generate a small gauge coupling.

There are, however, some problems with this argument:
\begin{itemize}
\item
In general, as noted by these authors, the scale of the ``low
energy theory," in these cases, is of order $M_p$, i.e.
$e^{-{2\pi \over b_a \alpha_a{T}}}\sim 1$.  These authors then
argue that this is a mechanism which can produce a small coupling,
but that because the scale is large, one must suppose that
supersymmetry is unbroken by this superpotential.  However, the
problem is deeper.  The low energy analysis is simply inconsistent
in this case.  There is no sense in which there is any
small parameter which might justify the analysis.  So, while this
may provide a toy model for how small couplings might be
generated, it does not explain why anything would be calculable,
and there is no hope that it corresponds to any sort of reliable
analysis one could hope to do in some limit of string theory.
\item
In general, it is not possible to argue that this is the correct
form of the superpotential at general points in the moduli space,
even for very small values of all of the couplings.  The problem
is that the usual sorts of symmetry arguments or dynamical
arguments which are made for the superpotential do not generalize
straightforwardly in the case of several gauge groups.
It is, however, possible to argue that the stationary points one
finds in this way may indeed be stationary points of the true low
energy effective action.  These issues will be discussed
elsewhere.
\end{itemize}

As noted by the authors of \cite{kl} (and was
an essential feature of the original racetrack
proposals), a discrete fine tuning can give a
hierarchical ratio of scales, and here we see that this is
essential if there is to be {\it any} sense in which quantities
may be calculated.  To simplify the notation, call $b_i = 3N_i$,
and suppose that there are two groups with very similar $\beta$
functions.  Then the stationary point for $S$
is
\beq
S= {N_1 N_2 \over N_1-N_2} \ln({C_2 N_1 \over C_1 N_2})
\eeq
If $N_1 \approx N_2$, this is of order $1/N^2$.\footnote{Note
that it is important that $C_1/C_2$ not be too close to
one, or $S$ will not be sufficiently large.  In the model
of ref. \cite{iy} which we will discuss below, the $C_i$'s
involve independent couplings, so there is no reason
that the ratio should be close to one.}  In this case,
the exponential is $e^{-1/N}$, and thus is hierarchically small.
As a result, the superpotential and gauge coupling functions
receive only exponentially small corrections to their values for
very large $S$.  So there is some hope that {\it holomorphic}
quantities are calculable.  Inherently stringy effects are of
order $e^{-N^2}$, so unless they have enormous coefficients,
they are negligible.

One can now ask:  are non-holomorphic quantities calculable?  One
might expect that the answer is no, for the following reason.
Suppose that one calculates, for example, corrections to the
Kahler potential for $S$.  In the heterotic string, these are down
by $g^2$, but there are loops with $N^2$ particles, and $g^2 \sim
N^2$.  This was one of the points made in \cite{dsracetrack}.
However, there are some possible loopholes to this argument.
First, other moduli may appear.  In the case of the heterotic
string, loop diagrams may involve extra factors of $1/T$.  To see
this, consider the low energy effective theory, and consider the
usual form of the weak coupling effective action.  In this action,
the kinetic term for $S$ has a factor of $1/S^2$ out front.  The
one loop term has two factors of $S^{-1}$ from propagators, but
also a factor of ${\rm cutoff}^2/M_p^2$.  Here the cutoff should
be the string scale, which is related to the Planck scale by a
factor of $S$.  So there is no factor of $T$ here.  But in the
strongly coupled heterotic theory, the story is different.  The
cutoff is presumably the eleven-dimensional scale, which is
related to the Planck scale by a factor of $ST$.  So one gets an
additional factor of $T^{-1}$ in the result.  So if $T$ is somehow
large as well, it is possible that the perturbation expansion
might make sense.

For the Type I theory, the situation is different again.  Here,
the loops contributing to the $S$ Kahler potential come with
factors of $g_s^2 N^2$, but $g_s \sim g_{YM}^2$, so now loops are
in some sense suppressed by $1/N^2$.  (I thank I. Antoniadis for
some remarks which prompted this discussion.)

Of course, in both the strongly coupled heterotic and the Type I
case, one does not know how to construct vacua with arbitrarily
large $N$, so it is not clear whether such a systematic
calculation is possible.  But perhaps we can be optimists.  After
all, we might imagine that for suitably small (but not
infinitesmal) couplings, the calculation of the low energy
effective action at the cutoff scale is reliable in string theory.
We might then expect that the low energy loop corrections are also
small for these values of the couplings.  So perhaps, under some
circumstances, the racetrack picture might provide a possibility
of understanding not only why couplings are small, but why
both holomorphic and non-holomorphic quantities might be
calculable.   Of course, there will be many other questions to
understand, most urgently the
cosmological constant.

Let us assume, for the moment, that only holomorphic quantites can
be reliably calculated.  Then we might want to limit our attention
to theories where supersymmetry is unbroken at the minimum, with
vanishing cosmological constant, which means
\beq
{\partial W \over \partial \phi} = W = 0.
\eeq
This can occur in theories with discrete $R$ symmetries.
A model (with continuous R symmetries, which can be generalized
trivially to the case of discrete R symmetries) was suggested
in \cite{iy}.  The model is somewhat complicated, requiring many
scalar fields.  Still, it provides an existence proof that such a
stabilization is possible, at least in principle.  In such a
theory, we might well hope to calculate holomorphic quantities.
For these, the expansion parameter would be $e^{-1/N}$, and this
is quite plausibly small enough.
Non-holomorphic quantities, we have seen, might also be
calculable, depending on whether $1/N$ can be thought of as a
small expansion parameter.  We have argued that this might be
wishful thinking, but it is not totally implausible.

If supersymmetry is broken at the minimum, there are many issues
which must be dealt with.  First, it is less clear that the
superpotential used at the minimum is reliable; one needs to rely
on $1/N$ as a small parameter to assess this.  Said another way,
if there are large corrections to the Kahler potential, they can
effect the location of the minimum.  One will have to face more
directly, in this case, the problem of the cosmological constant,
even in determining the location of the minimum.  Still, the fact
that, in some sense, even Kahler potential corrections {\it might}
be calculable in some circumstances is an intriguing one.

\section{Some Cosmological Questions}

There are a number of ways in which moduli are likely to be
relevant to cosmology.  One might imagine that they could act as
inflatons\cite{gaillard}; this view has been persuasively
put forth recently by Banks\cite{banksinflation}.
In the rest of this section, we pursue some of the cosmological
issues associated with the Kahler stabilization and racetrack
scenarios.

A number of cosmological difficulties related to moduli have been
considered in string theory.  One is the cosmological moduli
problem\cite{bkn}.  A number of solutions have been proposed to this
problem and I will not review them systematically here.  Let me mention
just two possibilities.  First, it could be that the moduli are
heavier than naively expected, i.e. $10$'s of TeV\cite{bkn}.  Then their decays
restart nucleosynthesis.  In general, one might worry that it is difficult
in such a scheme to produce baryons, but the authors of \cite{yanagidaetal}
have argued that in the presence of an Affleck-Dine condensate,
there is no problem.  An alternative is that the minima of the
moduli potential are at points of enhanced symmetry.  The usual
problem with this is that one might expect any
enhanced symmetry point for the dilaton (that modulus which determines
the values of the observed gauge couplings)to lie at $\alpha \sim 1$.
Some scenarios where this might not hold have been considered,
but the
racetrack picture suggests another possibility\cite{inpreparation}.  Perhaps
this coupling is fixed by this mechanism, in such a way that
supersymmetry is unbroken and this modulus is very massive.
The other moduli could then sit at enhanced symmetry points,
and the universe could naturally find themselves sitting
at such a point.

This then raises the question of whether the dilaton itself, even
if massive, is likely to settle into its minimum, and this is the
question I would like to focus on in the rest of this talk.
This problem was discussed by
Brustein and Steinhardt some time ago\cite{bs} (though I understand that
Peskin et al were aware of this issue).  They observed that
if the superpotential really has the structure suggested by gluino
condensation (say with stabilization as in the
racetrack scheme,
or through Kahler stabilization), it is difficult for the system
to find its true vacuum.  The point is readily illustrated if we
use the Kahler potential suggested by the weakly coupled theory.
Then the canonical field (say corresponding to $S$) has the form:
\beq
S = e^{\phi}
.
\eeq
So the potential behaves roughly as
\beq
e^{-1/N} e^{\phi}.
\eeq
This is extremely steep, and the difficulty is that the system is
likely, even if it starts at larger coupling than the coupling at
the minimum, to overshoot the minimum.

One might object that this argument relies on using the weak
coupling form of the Kahler potential, while we have argued that,
even though the superpotential might be similar to that expected
from the weak coupling analysis, the Kahler potential is likely to
receive large corrections.  Nir, Shadmi, Shirman and I have looked at
this problem, however, and found that modifications of the Kahler
potential can only solve the problem if one does drastic fine
tuning.  If one only fine tunes the Kahler potential and its first
and second derivatives near the minimum, this is not enough.
Essentially, the Kahler potential must be fine-tuned over a
finite range in field space around the minimum.

A second possible resolution of this problem is provided by
observations in \cite{banksetal}.  These authors note that it is
not consistent to suppose that zero-momentum moduli dominate the energy
density at early times.  The point is easy to understand.
Consider the case of the weak coupling dilaton again.  Assume that the
energy density density is dominated by $\phi$. In any
regime where the potential can be neglected, the equation of state
is simply $p = \rho$, so the scale factor grows as
$R(t) \sim t^{1/3}.$ As a result, the canonical field
obeys the equation:
\beq
\ddot \phi + {1 \over t} \dot  \phi =0
,\eeq
with solution
\beq
\dot \phi = {c \over t}~~~~~~ \phi = \ln t + d.
\eeq
From this equation, it is clear that the potential falls off much
more rapidly with time than the kinetic terms; this is the
Brustein-Steinhardt problem again.   Now consider, however, the
equation for the non-zero momentum modes.  This is
\beq
\ddot \phi  + {1 \over t} \dot \phi + {k^2 \over R^2} \phi =0
\eeq
with solution
\beq
\phi = {1 \over t^{1/3}} \cos{(3/2) t^{2/3}}.
\eeq
As a result, the energy density of the non-zero modes falls off as
$t^{4/3} \sim R^4$, i.e. more slowly than that of the zero modes
(which falls off as $1/t^2$), and just like that of radiation.

\begin{figure}[htbp]
\centering
\centerline{\psfig{file=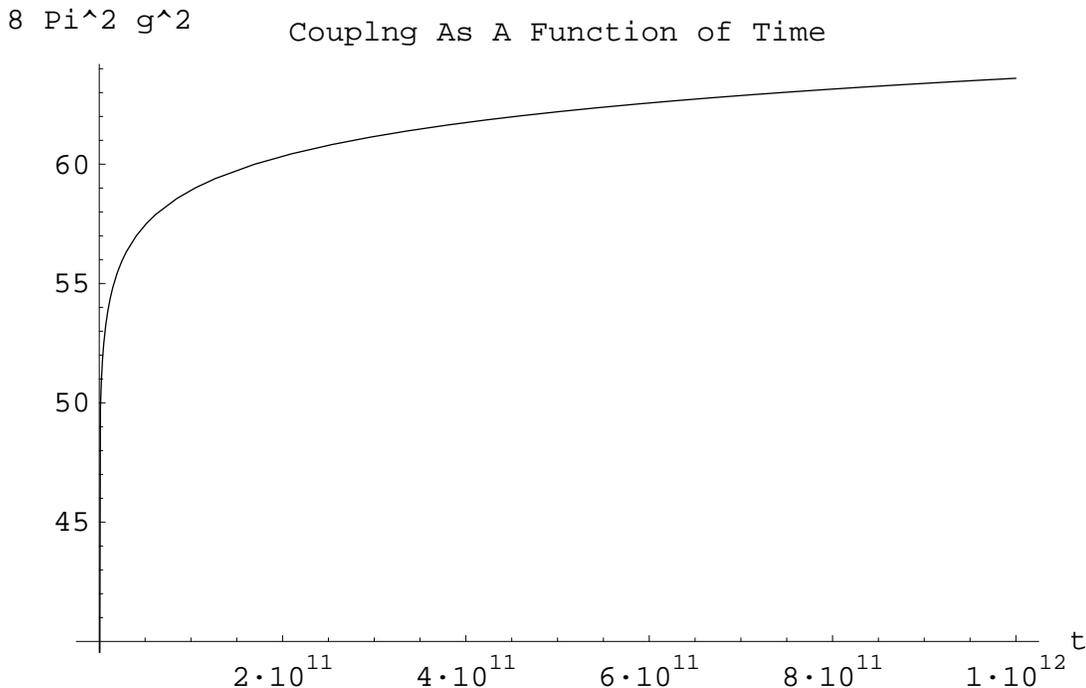,angle=-0,width=15cm}}
\caption{Coupling as a function of time in a radiation dominated
universe.}
\label{radiationdominated}
\end{figure}

\begin{figure}[htbp]
\centering
\centerline{\psfig{file=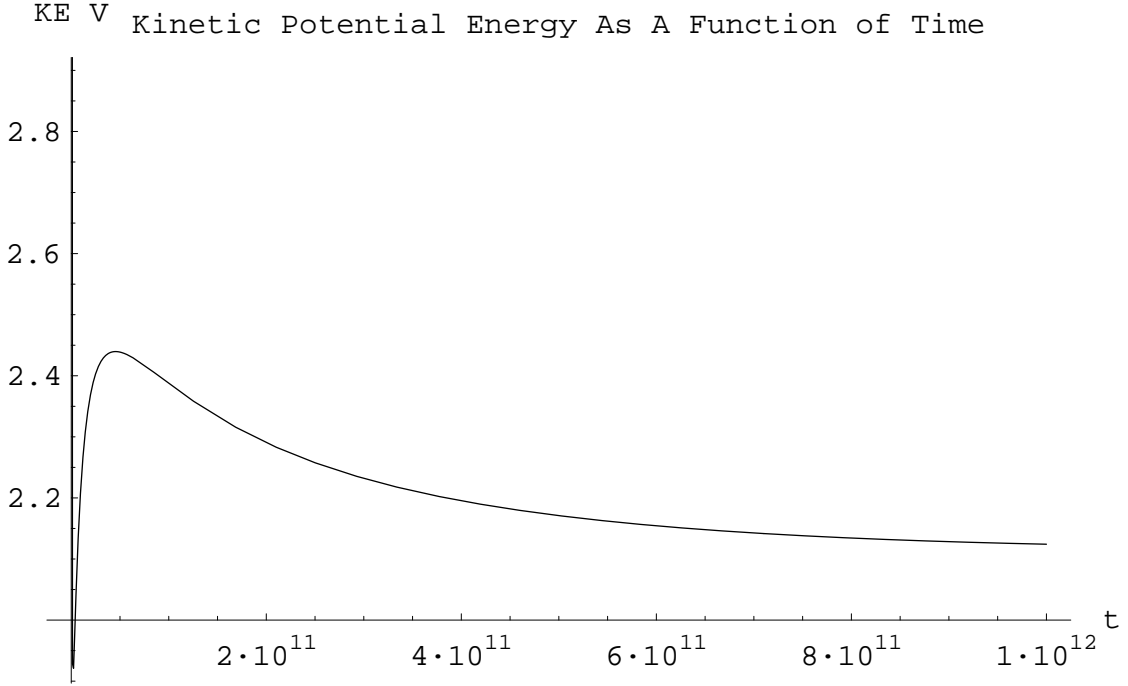,angle=-0,width= 15cm}}
\caption{Ratio of kinetic to potential energy.}
\label{kineticenergy}
\end{figure}

So, while we do not really understand what might be appropriate
initial conditions for this system, it is clear that assuming that
the field is homogeneous on some scale is not self consistent.
Because non-zero momentum modes behave like radiation (e.g.
assuming isotropy, $p = 1/3 \rho$), one way to model this system
is by supposing that the universe is radiation dominated.  Note that
this does
not require that the system be in thermal equilibrium; simply that
the non-zero modes of the moduli dominate the energy density, and are
roughly isotropic.
In this case, the equations of the problem are different.  In
particular, if one can neglect the potential, the zero mode now
obeys the equation
\beq
\ddot \phi + {3 \over 2t} \dot \phi =0.
\eeq
This has solution
\beq
\phi = a t^{-1/2} + b.
\eeq
In other words, the field creeps to some particular point.
Including the potential, it is now reasonable to hope that the
system will track the potential, and eventually settle into the
correct minima.  Numerical study indicates that this does indeed
occur for a range of initial conditions.  This can be seen in
figs. [1,2].   In these figures,
the evolution of the coupling is plotted in a purely exponential
potential.  One sees that the motion of the field is very slow;
moreover, its kinetic energy is never much larger than its
potential energy.   As a result, when the field reaches the
minimum, it does not overshoot.

There is, however, a difficulty with this picture.  While the
energy may be dominated by the kinetic terms of some field(s),
these themselves will receive $g^2$-dependent corrections. These
corrections may well be much larger at the relevant times then the
non-perturbative corrections to the potential, and may shift the
location of the minimum.  In this case, one must make some
assumption about these corrections and follow more carefully the
evolution. In particular, the potential might well force the
system to weak coupling and away from the desired minimum. In this
case, the Brustein-Steinhardt problem
is replaced by an even more severe
difficulty.  On the other hand, if the potential
has a local minimum at relatively strong coupling,
a gentle landing is still possible.
This problem will be described in detail elsewhere, but its essential
features are modeled by the situation where the system is truly in
thermal equilibrium, which we turn to now.

An alternative possibility is that the system is truly in thermal
equilibrium, i.e. that the gauge bosons, etc., are all in thermal
equilibrium.  In this case, the analysis required is quite
different.   For all but very large values of the coupling, the
zero temperature dilaton potential is irrelevant; the largest
contribution comes from the coupling-dependence of the free
energy.  In particular, provided $T \gg \Lambda(S)$, we can simply
take over the high temperature expression for the free energy.
This has the form
\beq
V(g^2,T) = T^4(-a + b g^2 + c g^3 + \dots).
\eeq
If one examines the explicit form for the coefficients $a,b,c$ in
the case of $SU(N)$\cite{kapusta}, one
finds that for  $N=10$ the coefficient $c$ is greater than $b$ for
$g \sim 1$, so
the perturbation expansion is already not reliable for rather
modest $N$.

One can then
imagine several possible behaviors for the effective
potential as a function of $g$.  It might have no minimum, simply
falling to zero at small coupling.  In this case, the system
will not find the true minimum.  It might have a local minimum at
some $g= g_o(T)$, where given that the potential
for large coupling ($\Lambda > T$) grows so steeply,
 $g_o(T)$ decreases with temperature.  In this case, one might
 expect that the system will roughly track the minimum,
 and that it will end up being set rather gently in the true
 minimum.  We have checked that this does occur for a range of
 initial conditions.

To summarize, the extent to which the Brustein-Steinhardt problem
is a problems
seems quite sensitive to the initial conditions, which are
certainly not well understood at present.

\section{Conclusions}

The question ``what is string theory," has been given new focus by
recent proposed solutions to the hierarchy problem.  Susskind has
argued that the only aspects of this theory which we really
understand are those associated with states with a high degree of
supersymmetry.  This is troubling, as he notes,
because these states do not resemble the world we observe.
I have argued that we can, with some
confidence, extend this list at least to certain exact moduli
spaces with four supersymmetries, and to {\it approximate moduli
spaces} which, in certain limits, have four supersymmetries.  Two
facts of nature suggest that the world we see might
sit on such a space:  the
smallness of the gauge couplings and the existence of hierarchies.
While generically one might expect that couplings should be of
order one and all scales in the theory should be comparable,
we have reviewed at least two plausible mechanisms for fixing the
moduli on such a space.  We have seen that some -- and conceivably
all -- quantities might be calculable in this circumstance.

This discussion has also allowed us to frame certain issues
related to proposals for a low string scale as a solution to the
hierarchy problem.  We have asked what it would
mean for string theory to predict low energy supersymmetry, and
argued that this would correspond to demonstrating that the ground state
of the theory should lie on an approximate moduli space with N=1
supersymmetry in the sense described above.  The low
string scale proposals are generally not in this class.
In
order to explain a large hierarchy of scales, most of these
proposals invoke supersymmetry in the bulk.
But exact supersymmetry is recovered, if at all, only in the limit where the
extra dimensions become infinitely large, so that (from a
four-dimensional perspective) there are an infinite number of
gravitinos.  Interestingly, the proposal of \cite{ibanez} has a
low energy phenomenology much like that of the N=1 models.  It has
been argued\cite{goldbergerwise} that in the Randall-Sundrum
picture, supersymmetry might not be necessary even in the bulk to
understand the hierarchy.  It would be interesting to investigate
this statement carefully, and in particular to make sure that the
results found in a classical analysis are not spoiled by quantum
effects.

The real question, however, is whether there is any way to
understand why approximate moduli spaces with low energy
supersymmetry should or should not, in any sense
be favored by string dynamics.
There has long been a view that this is the case, but it is
largely based on prejudice.  Settling this question could be the
basis of a {\it prediction} of low energy supersymmetry from
string theory, or alternatively of large dimensions, with their
distinctive consequences for future experiments.

Within the framework of supersymmetric approximate moduli, we have
also discussed some cosmological issues.  We have seen that the
most often discussed
problems of string cosmology might be solved if the dilaton -- the
field which fixes the ordinary gauge couplings -- is fixed at a
large scale, without breaking supersymmetry, while the other
moduli sit at enhanced symmetry points.  Such a picture makes
distinct low energy predictions.  It predicts, as has been
discussed elsewhere\cite{dns,yukawa} low energy supersymmetry {\it
and} low energy breaking of supersymmetry.


\end{document}